# Observation and Spectroscopy of a Two-Electron Wigner Molecule in an Ultra-Clean Carbon Nanotube

S. Pecker*[1], F. Kuemmeth*[2], A. Secchi[3,4,†], M. Rontani[3], D. C. Ralph[5,6], P. L. McEuen[5,6] and S. Ilani[1]

[1]*Department of Condensed Matter Physics, Weizmann Institute of Science, Rehovot 76100, Israel*

[2]*Center for Quantum Devices, Niels Bohr Institute, University of Copenhagen, Universitetsparken 5, DK-2100 Copenhagen, Denmark*

[3]*CNR–NANO Research Center S3, Via Campi 213/a, 41125 Modena, Italy*

[4]*Department of Physics, University of Modena and Reggio Emilia, 41125 Modena, Italy*

[5]*Physics Department, Cornell University, Ithaca, New York 14853, USA*

[6]*Kavli Institute at Cornell, Cornell University, Ithaca, New York 14853, USA*

[†]*Present address: Institute for Molecules and Materials, Radboud University of Nijmegen, 6525 AJ Nijmegen, The Netherlands*

**Coulomb interactions can have a decisive effect on the ground state of electronic systems. The simplest system in which interactions can play an interesting role is that of two electrons on a string. In the presence of strong interactions the two electrons are predicted to form a Wigner molecule, separating to the ends of the string due to their mutual repulsion. This spatial structure is believed to be clearly imprinted on the energy spectrum, yet to date a direct measurement of such a spectrum in a controllable one-dimensional setting is still missing. Here we use an ultra-clean suspended carbon nanotube to realize this system in a tunable potential. Using tunneling spectroscopy we measure the excitation spectra of two interacting carriers, electrons or holes, and identify seven low-energy states characterized by their spin and isospin quantum numbers. These states fall into two multiplets according to their exchange symmetries. The formation of a strongly-interacting Wigner molecule is evident from the small energy splitting measured between the two multiplets, that is quenched by an order of magnitude compared to the non-interacting value. Our ability to tune the two-electron state in space and to study it for both electrons and holes provides an unambiguous demonstration of the fundamental Wigner molecule state.**

Two electrons confined to a one-dimensional string form one of the simplest realizations of an interacting quantum-mechanical system. The behavior of this system is governed by the balance between kinetic and interaction energies; When kinetic energy dominates, the electrons occupy particle-in-a-box levels along the string. In contrast, when interactions dominate, a Wigner-molecule ground state is formed, in which the repulsion of the two electrons drives them to localize at the two sides of the string[1,2]. Owing to the fermionic nature of the two particles their total wavefunction is anti-symmetric with respect to electron exchange, leading to an intimate connection between their real space and spin-space behaviors. Consequently, the real-space charge separation in a Wigner molecule goes hand in hand with a spin-space signature, namely a dramatic quenching of its spin excitation energies[3].

Carbon nanotubes (NT) are an excellent system to search for the existence of a Wigner-molecule ground state. This system is known to have strong electron-electron interactions[4–8], and due to recent technological breakthroughs[9] it is now possible to fabricate NT devices clean enough to allow measurements down to the single-carrier limit[10–12]. Compared to III-V semiconductor systems[13–16] in which Wigner molecule formation has been explored previously, in suspended NTs the screening of Coulomb interactions is strongly reduced and the one-dimensional confinement potential for electrons or holes can be shaped with gate electrodes. This ability to control the confining potential is critical because it allows one to distinguish between extrinsic electrostatic effects that spatially separate the two electrons and intrinsic separation driven by their repulsion. Furthermore, in addition to the conventional two-fold spin degeneracy in other semiconductors, electrons in NTs possess a two-fold orbital degeneracy (isospin), forming a four-fold spin-isospin subspace. Recent experiments[10] have shown that the electrons' spin and isospin in NTs are easily polarized by magnetic fields, which has been interpreted as an indication of Wigner-crystal order. However, more recently single-particle spin-orbit coupling has been found in this system[11,17,18], which similarly to interactions can also preferentially align the spins and isospins. In an attempt to unambiguously identify the effects of interactions, recent theoretical works have focused on the case of two electrons, and have demonstrated that the role of interactions can be directly determined by measuring the excitation spectrum[19–23]. Two-electron excitations also play a key role in making quantum bits in NTs[24]. However, since interactions may hinder qubit implementation by suppressing Pauli-blockade physics[12,19,23], experiments so far have been done in the non-interacting regime (due to geometry, dielectric environment, and level spacing)[18,25–28]. A measurement of the excitation spectrum of two electrons in the opposite regime of strong interactions has so far been missing, and holds the key for determining the strongly-interacting nature of this system.

In this work we probe the excitation spectrum of two carriers, electrons or holes, confined to a NT quantum-dot by transport spectroscopy. We identify seven low-energy quantum states that fall into two multiplets that are symmetric or anti-symmetric under particle exchange in real space. We find that a single-particle description of the two-electron system with spin-orbit coupling captures well the energy spectrum within each multiplet. Interestingly, however, the energy splitting between the two multiplets is quenched by an order of magnitude compared to its non-interacting value. We show that this quenching is a direct manifestation of the formation of a Wigner-molecule state. By measuring similar spectra for electrons and holes, having opposite response to disorder potential, we demonstrate the generality of our observation and the irrelevance of disorder.

Our device, used previously to study spin-orbit coupling of one electron (1$e$) in a single quantum dot, is now used to study two-electron (2$e$) states in a molecular regime. We obtain essentially identical results for two holes (Supplementary S5). The device consists of a NT suspended above a pair of split-gate electrodes and contacted by source and drain electrodes (Fig. 1a). The charge stability diagram, measured as a function of the common voltage on the gates, $V_g$, and their difference (detuning), $\epsilon$, (Fig. 1b) shows a rounded honeycomb structure, similar to that of a strongly tunnel-coupled double dot. This molecular configuration allows us to continuously transform between two different 2$e$ configurations: In one, the electrons are localized in different sites near the two ends of the NT (the (1,1) configuration), whereas in the other, both electrons occupy the same site (the (0,2) configuration). Figures 1c and 1d show the measurement of the 2*e* excitation spectra in these two configurations. For each configuration we measure the conductance as a function of source-drain bias, $V_{sd}$, and $V_g$. The parallel lines observed within these "Coulomb diamonds" correspond to the individual 2*e* excited states. Below we will show that the different excitation spectra, observed in these two configurations that differ by the detuning, provide crucial information for understanding the role of interactions in these molecular states.

The interacting nature of two-electron states is expected to be clearly imprinted on their detuning-dependent energy spectrum, shown schematically for the non-interacting case in Fig. 2a, and for the strongly-interacting Wigner-molecule case in Fig. 2b. On the left side of the figure (low detuning) the two electrons are in the (1,1) molecular configuration, separated to the two sides of the NT. On the right side, the two electrons are in (0,2) molecular configuration, both occupying the same side. Since detuning increases the energy of the left side with respect to that of the right side, each state in the (1,1) configuration rises in energy with detuning, whereas each state in the (0,2) configuration falls in energy. In the figure we color the 2*e* states according to their symmetry with respect to electron exchange in real space: The ground state is always symmetric (S) in real space (red) whereas the first excited state is anti-symmetric (AS) in real space (blue). In the non-interacting limit and when the two electrons are on the same side (right side of Fig. 2a) in a spatially symmetric 2*e* wavefunction the two electrons can both occupy the lowest particle-in-a-box level, whereas in a spatially anti-symmetric wavefunction one electron must occupy the next particle-in-a-box level. Thus, in the non-interacting limit there is a large symmetric – anti-symmetric splitting, $\Delta_{S-AS}$, equal to the single-particle level spacing, $\Delta_{ls}$. The situation is very different in the presence of strong interactions (Fig. 2b), which drive the two electrons apart in real space. Such electronic separation has very little effect on the anti-symmetric state, where the electrons are already separated in real space by virtue of symmetry, but it has a dramatic effect on the symmetric ground-state, in which in the absence of interactions both electrons strongly overlap. Interactions drive the density profile of the symmetric and antisymmetric states to be essentially identical (Fig. 2b), and correspondingly, their energy splitting becomes strongly suppressed, $\Delta_{S-AS} \ll \Delta_{ls}$. The symmetric – anti-symmetric splitting, $\Delta_{S-AS}$, thus serves as a quantitative measure of the interaction strength and the real-space separation of a Wigner molecule.

The same reasoning holds regardless of whether the two electrons occupy the (1,1) or (0,2) configuration. In fact, in the (1,1) case the suppression of $\Delta_{S-AS}$ should be even more dramatic than for the (0,2) case. As will be shown below we indeed measure, as in previous works[13,16] suppressed $\Delta_{S-AS}$ for the (1,1) configuration. We note, however, that in the (1,1) configuration each electron is near an edge, and as such it is inherently attracted to its image charge in the metallic contact, effectively creating an artificial double-dot potential. This effect is as large as the repulsion between the electrons themselves and would act similarly to separate them in space. To critically test the effect of interactions between the electrons we must therefore localize them near one edge, effectively in a single dot. Then the attraction to an image charge on the same edge has an *opposite* effect than the electronic repulsion. The ability to squeeze the electrons to one side of the tube, used in this work, is thus fundamental for pinpointing the effects of their mutual repulsion.

To adapt the above picture to a NT we need to recall that for each particle-in-a-box state along the NT there are four possible spin-isospin combinations, whose degeneracy is broken by spin-orbit interactions[11]. Correspondingly, states of two electrons have $4 \times 4 = 16$ possible spin-isospin combinations, 6 of them are spatially symmetric and 10 are spatially anti-symmetric (Supplementary S1). Thus each line in the schematic diagrams of Figs. 2a or 2b should appear in multiple copies. However, only some copies should be visible in transport experiments that probe 1*e*-2*e* transitions, as illustrated in Figs. 2c and 2d. In all cases the system starts with one electron in the lowest single-particle state (gray symbol) and a second electron hops in and out, providing a conductance signal. When the two electrons form a symmetric state in real space they cannot occupy the same spin-isospin state (Supplementary S1), leaving only the three high-lying states for the added electron (Fig. 2d). When they form an anti-symmetric state in real space all four states are available (Fig. 2c). Thus in total, seven out of the sixteen 2*e* states should appear in transport.

A clear way to experimentally distinguish the symmetric and anti-symmetric multiplets is by their magnetic-field fingerprints. A magnetic field parallel to the tube axis, $B_\parallel$, couples to both the orbital and spin magnetic moments of the electron (up and down arrows in Figs. 2e-f) and shifts the energy of each spin-isospin state with a unique slope. As was previously demonstrated[11], for a single electron this results in a characteristic "double-cross" structure, split at $B_\parallel = 0$ by spin-orbit coupling, $\Delta_{SO}$. For two electrons in the non-interacting framework the addition spectrum measured by transport amounts to the energy of the added electron, resulting again in simple fingerprints: In the 2e anti-symmetric multiplet the added electron can populate all four spin-isospin states and thus the corresponding addition energies will have a double-cross pattern (Fig. 2e) identical to that of a single electron. In the 2e symmetric multiplet the lowest state is forbidden, leading to a double-cross without the lowest line (Fig. 2f), having a distinctive cusp at $B_\parallel = 0$. As a function of detuning these multiplets should evolve, depending on the strength of interactions, either like the non-interacting case (Fig. 2a) or as the strongly interacting case (Fig. 2b), only that now even at $B_\parallel = 0$ we should see two copies of each line, split by $\Delta_{SO}$ (Fig. 2g). Using the above identification tools we can proceed to experimentally study the 2e excitations. We start by measuring their dependence on $B_\parallel$ at three different detunings: in the (1,1) configuration (Fig. 3a), the (0,2) configuration (Fig. 3c), and at the crossover between them (Fig. 3b). In all figures we plot the conductance, measured at a finite $V_{sd}$, as a function of $B_\parallel$ and $V_g$ (converted to energy on the right axis). Each line of enhanced conductance arises from a 2e state, giving directly the magnetic-field dependence of the 2e energy spectrum. Looking first at the (1,1) configuration (Fig. 3a) we identify four lines, two with positive slopes and two with negative slopes. Notably, the magnetic moments and the zero-field splitting ($0.34 \pm 0.01$ meV at $B_\parallel = 0$) in this 2e spectrum are identical to those in the one-electron double-cross spectrum we reported earlier[11] showing that the observed splitting is due to spin-orbit coupling (Supplementary S3). This allows us to clearly identify the spin and isospin quantum numbers of each state. At higher detuning (Fig. 3b) we observe additional excitations, most apparent as a cusp at $B_\parallel = 0$. Finally at even higher detuning in the (0,2) configuration (Fig. 3c), the double-cross (visible at ~1 meV above the ground state) remains the strongest feature, while the additional multiplet with a zero-field cusp fully emerges at low energies.

The measured magnetic fingerprints (Figs. 3a-c) show remarkable similarity to the predicted ones (Figs. 3d-f) based on the non-interacting theory (Figs. 2e-f). The low-energy, cusped multiplet observed at high detuning (Fig. 3c) can thus be identified with the symmetric multiplet (red lines, Fig. 3f), and the double-cross at higher energies with the anti-symmetric multiplet (blue lines, Fig. 3f). Experimentally, we observe one additional cross between the multiplets (Fig. 3c, at ~0.6 meV above the ground state), which we associate with inter-valley backscattering processes (Supplementary S4). However, apart from it, the non-interacting framework quantitatively describes the entire structure within each multiplet. With decreasing detuning, the splitting between the multiplets decreases (Fig. 3b), until they fully overlap in the (1,1) configuration (Fig. 3a). We observe a spectrum identical in all of the above details for two holes (Supplementary S5), demonstrating that all these observations are generic.

The crucial role played by interactions is unraveled when we measure the detuning dependence of the excitations (Fig. 4a). This figure plots the conductance at $B_\parallel = 0$ as a function of $\epsilon$ and $V_g$ (converted to energy on the right y-axis). The lowest line corresponds to the $2e$ ground state. A parallel line, at energy $eV_{sd}$ above it (labeled $W$), arises from the width of our spectroscopic window set by $V_{sd} = -2$ mV. The lines in between are the $2e$ excitations. These excitations are visible in this figure mostly at high detuning due to asymmetric coupling to the leads, while measurements at opposite $V_{sd}$ (Supplementary S6) reveal their complementary dependence at low detuning, and the dashed guiding lines fit the data from both. By associating each excitation with its magnetic fingerprint in Fig. 3 we identify the pair of symmetric lines (red dashed), the pair of anti-symmetric lines (blue dashed) and the single line in between (due to inter-valley backscattering, Supplementary S4). Notably *all* the excitation lines evolve from an up-going slope at low detuning to a down-going slope at high detuning, clearly indicating that all of them completely evolve from the (1,1) to the (0,2) configuration. The spectrum measured at high detuning shown in Fig. 3c is thus a direct observations of the energy spectrum of two electrons in a *single* dot (similarly for two holes in Supplementary Fig.

S4). As explained above, in the absence of interactions, the splitting between symmetric and anti-symmetric states should amount to the single-particle level spacing, $\Delta_{ls}$. In Supplementary S2 we directly extract this single-particle spacing from measurements of the 1$e$ spectrum to be $\Delta_{ls} = 7.8 \pm 0.1$ meV. Remarkably, if we compare this spacing with the symmetric to anti-symmetric splitting in the 2$e$ spectra we see that this 2$e$ excitation energy is quenched by almost an order of magnitude compared to its non-interacting value.

The drastic quenching of the symmetric – anti-symmetric excitation energy as compared to the non-interacting picture attests to the effect of strong electron-electron interactions. To better understand the role of interactions we performed an exact-diagonalization calculation[29] of the excitation spectrum (Supplementary S7) corresponding to the parameters of our 2$e$ dot, as a function of the dimensionless interaction strength $r_s = d/a_B^*$, where $a_B^*$ is the effective Bohr radius and $d$ is the length scale of the harmonic oscillator potential. Without interactions ($r_s = 0$) the ground state is of the symmetric multiplet, and the anti-symmetric multiplet is higher in energy by $\Delta_{ls} = 7.8$ meV. With increasing $r_s$, the anti-symmetric states drop in energy, becoming degenerate with the symmetric states for large $r_s$. In this limit the two electrons form a Wigner molecule, the transition to this state being continuous due to the one-dimensionality and small number of electrons. Our experimental observation of a ten-fold quenching corresponds to the $r_s = 1.64$ case (arrow, Fig. 4b). As explained above, the quenching follows from the spatial separation of the two electrons. This becomes apparent by comparing the calculated electronic charge-density profiles along the NT in both multiplets for the non-interacting (Fig. 4c) and interacting (Fig. 4d) cases. Indeed, the density profiles calculated for the observed quenching are nearly identical for the two multiplets, demonstrating the strongly-interacting nature of this two-electron Wigner molecule.

In summary, using transport spectroscopy of ultra-clean NT quantum dots we measure directly the excitation spectrum of two interacting electrons or holes. By tuning the 1D confinement potential we go between a state where the electrons are artificially separated by the confining potential to one where their separation is determined solely by their interactions. In the latter case we observe seven quantum states, grouped into two multiplets according to spin-isospin symmetry. The magnetic fingerprint within each multiplet is reproduced by the non-interacting picture. Remarkably, however, the fundamental excitation involving a change in symmetry is dramatically quenched in energy compared to its non-interacting value. Using exact-diagonalization calculations we demonstrate that such quenching is a fundamental signature of a strongly-interacting Wigner molecule, in which electrons are spatially separated by their mutual repulsion. The spectroscopy of the NT Wigner-molecule, provided here for the first time, directly shows that suspended carbon NTs can host strongly-interacting ground states and opens the way for studies of a wider variety of strongly-interacting multi-electron states predicted to exist in one-dimensional systems.

**Methods**

Devices were fabricated from degenerately doped silicon-on-insulator wafers, with a 1.5-µm-thick device layer on top of a 2-µm buried oxide. Using dry etching and thermal oxidation (thickness 100 nm) we isolated two electrically-independent mesas from the device layer that served as bottom gates to the NT. Gate contacts (2/50 nm Ti/Pt), source and drain electrodes (5/25 nm Cr/Pt) and catalyst pads were patterned using electron-beam lithography. Nanotubes were grown after completing all patterning to produce clean devices. All measurements were performed in a dilution refrigerator with a base temperature of $T = 30$ mK. The electron temperature extracted from the width of the Coulomb peaks was $100 - 200$ mK. The conductance was measured using standard lock-in techniques with small excitations (typically $4 - 10$ $\mu$V).

**Acknowledgements:** We would like to acknowledge E. Berg, A. Stern, A. Yacoby, and B. Wunsch for useful discussions. S.I. acknowledges the financial support by the ISF Legacy Heritage foundation, the Bi-National science foundation (BSF), the Minerva foundation, the ERC starters grant, the Marie Curie People grant (IRG), and the Alon fellowship. S.I. is incumbent of the William Z. and Eda Bess Novick career development chair. P.L.M acknowledges support by the NSF through the Center for Nanoscale systems, and by the MARCO Focused Research Center on Materials, Structures and Devices. The experiments used the facilities of the Cornell node of the National Nanotechnology Infrastructure Network (EECS-0335765) and the Cornell Center for Materials Research (DMR-1120296), both funded by NSF. M.R. and A.S. acknowledge support from Fondazione Cassa di Risparmio di Modena through the project COLDandFEW, from EU through the Marie Curie ITN INDEX, and from the CINECA-ISCRA supercomputing grant IscrC_TUN1DFEW.

**Figure 1: Excitation spectra of a two-electron molecular state in an ultra-clean carbon nanotube (NT). a**, Device schematic. A single ultra-clean NT is contacted by source and drain electrodes separated by 500 nm and suspended over two gate electrodes. The two gates induce a controllable electrostatic potential along the tube, as depicted by the energy band diagram. **b**, Measured charge stability diagram of the device. The differential conductance of the device, $G = \frac{dI}{dV_{sd}}$ (where $I$ is the current and $V_{sd}$ is the source-drain bias), is plotted as a function of common voltage $V_g = \frac{V_{gr}+2.4V_{gl}}{3.4}$ and detuning $\epsilon = \frac{V_{gr}-V_{gl}}{3.4}$ ($V_{gr}$ and $V_{gl}$ are the right and left gate voltages). Index pairs $(n,m)$ denote the charge configuration, where $n$ ($m$) is the number of electrons on the left (right) side of the NT. **c,** Conductance through the NT measured as a function of $V_g$ and $V_{sd}$, around the transition between (0,1) and (1,1) configurations (circle in panel b). The two parallel lines on the top left correspond to the *2e* states and the one on the right, marked W, corresponds to the edge of the spectroscopic window. **d,** Similar measurement around the transition between (0,1) and (0,2) configurations (star in panel b). More *2e* states are observed in this case as compared to panel c.

**Figure 2: Energy spectra predicted for two non-interacting and two strongly-interacting electrons in a NT. a**, Addition energy of the two lowest *2e* states as a function of detuning in the non-interacting limit. On the left side the electrons populate the entire NT ((1,1) configuration) and on the right they are localized on one side ((0,2) configuration). Colors correspond to the symmetry of *2e* states in real space (see labels). On the (0,2) side the splitting of these states, $\Delta_{S-AS}$, is equal to the single-particle level spacing, $\Delta_{ls}$. Side insets: charge density profiles calculated for the *2e* states in the (0,2) configuration. **b**, Same, but for the strongly interacting (Wigner molecule) limit **c**, Spin-isospin states contributing to the transport around the 1*e* to 2*e* Coulomb blockade transition, for a spatially anti-symmetric 2*e* state. The gray symbol is the starting 1*e* state and the blue symbols are the possible states for the added electron. Spin and isospin are denoted by the thin and thick arrows. **d**, Same for spatially symmetric 2*e* states. **e**, Addition energies of the spatially anti-symmetric 2*e* multiplet as a function of magnetic field parallel to the tube axis, $B_\parallel$. The resulting magnetic fingerprint features a double-

cross pattern, split at $B_\parallel = 0$ by the spin-orbit coupling, $\Delta_{so}$. **f**, Same, but for the symmetric multiplet. In this case a characteristic cusp is visible at $B_\parallel = 0$. **g**, The expected spectrum of 2e states in a NT at $B_\parallel = 0$, which should be similar either to that in panel a or to that in panel b, depending on the strength of interactions, but should also have two copies of each line due to the spin-isospin degrees of freedom. Numbers above the lines denote their $B_\parallel = 0$ degeneracy.

**Figure 3: Magnetic-dependent spectra of the two-electron molecular states at different detunings. a-c**, $G$ measured as a function of $B_\parallel$ and $V_g$ (converted to energy on the right axis) around the *1e* to *2e* transition. Lines of enhanced conductance correspond to tunneling via *2e* states. Panel a is measured in the (1,1) configuration ($\epsilon = 35$mV, $V_{sd} = 2$mV), panel c is in the (0,2) configuration ($\epsilon = 83$mV, $V_{sd} = -1.7$mV) and panel b corresponds to a detuning in between these configurations ($\epsilon = 65$mV, $V_{sd} = -7$mV) . The line labeled W in panel c corresponds to a similarly labeled line, e.g. in Fig. 1d, representing the edge of the spectroscopic measurement window. In panel c we use the symbols from Figs. 2e-f to specify the spin and isospin content of each *2e* state. **d-f**, Theoretically calculated addition energies as a function of magnetic field for the detunings in panels a-c. Energies of the spatially symmetric states (red) and spatially anti-symmetric states (blue) are calculated with the measured spin and orbital magnetic moments. The splitting between the symmetric and anti-symmetric multiplets in each panel is chosen to fit the measurement.

**Figure 4: Quenching of excitation energies in a Wigner molecule. a**, $G$ measured as a function of $V_g$ and $\epsilon$ at $B_\parallel = 0$ and $V_{sd} = -2$ mV, giving the detuning dependence of the NT *2e* states. To enhance the visibility of individual states we plot the $G$ normalized by its maximal value for every $\epsilon$. Dashed lines: guidelines marking the position of the anti-symmetric states (blue) and symmetric states (red), based also on a measurement at an opposite bias (Supplementary S6). Full lines: The non-interacting prediction for the energy of the anti-symmetric states (see Fig. 2a). $\Delta_{S-AS}^{(0,2)}$ labels the splitting between the symmetric and anti-symmetric multiplets at large detuning (accurately extracted from Fig. 3c). The *W*-label line gives the edge of the spectroscopic window. **b**, Exact diagonalization calculation of the excitation energies of the symmetric multiplet (red) and anti-symmetric multiplet (blue) in the (0,2) charge configuration as a function of interaction parameter $r_s$. An arrow marks the spectrum that best fits the experimental data ($r_s = 1.64$). States which do not appear in transport are omitted. **c**, Exact diagonalization calculation of the charge density as a function of position along the tube $x$, $\rho(x)$, in the ground state of the symmetric (red) and anti-symmetric (blue) multiplets for $r_s = 0$. **d**, Same, but for $r_s = 1.64$.

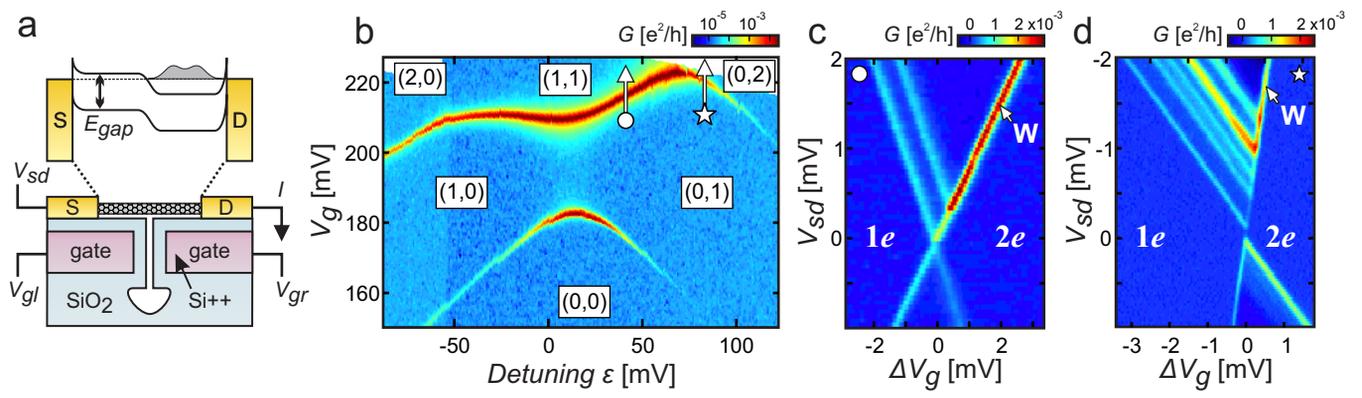

figure 1

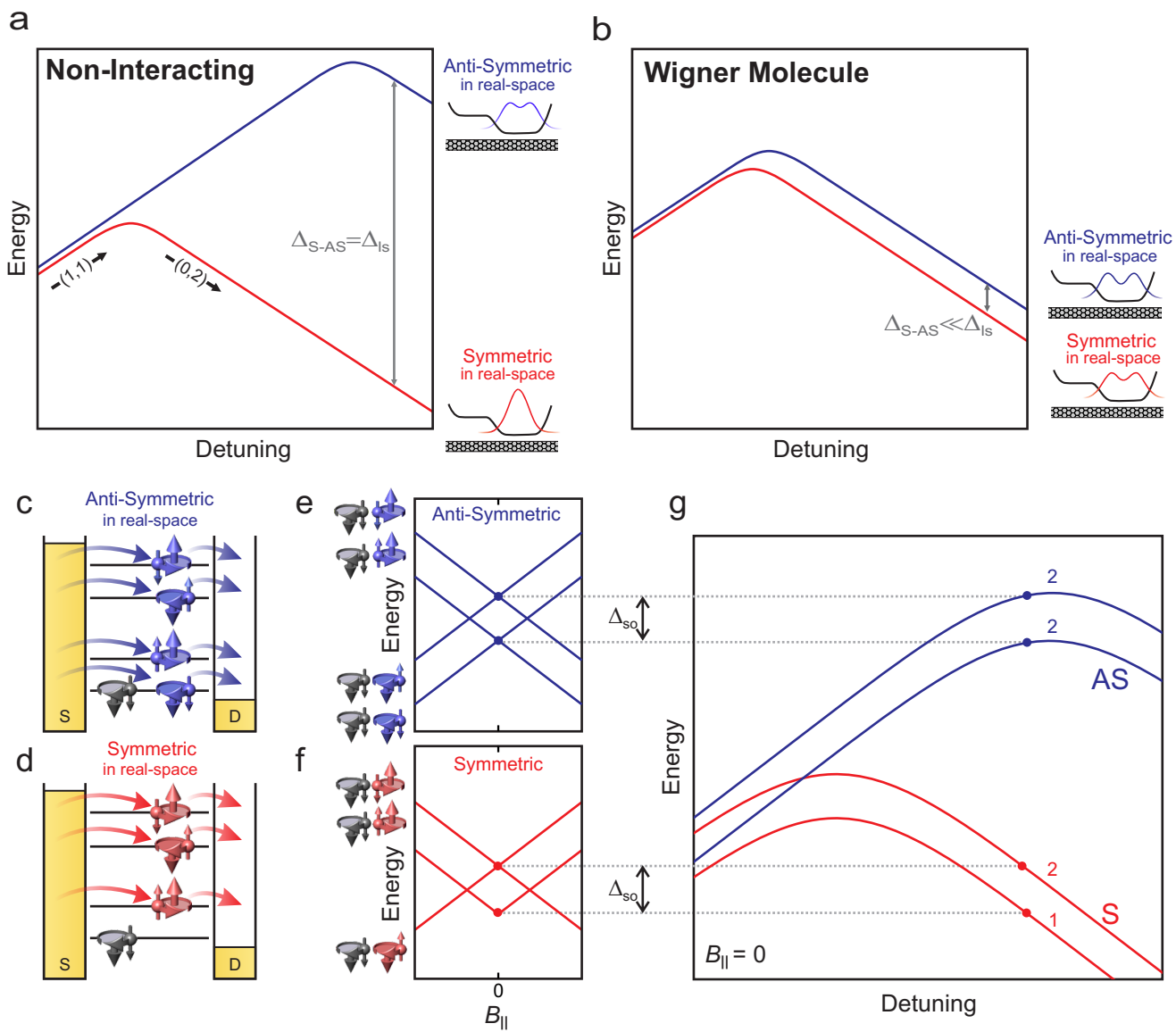

figure 2

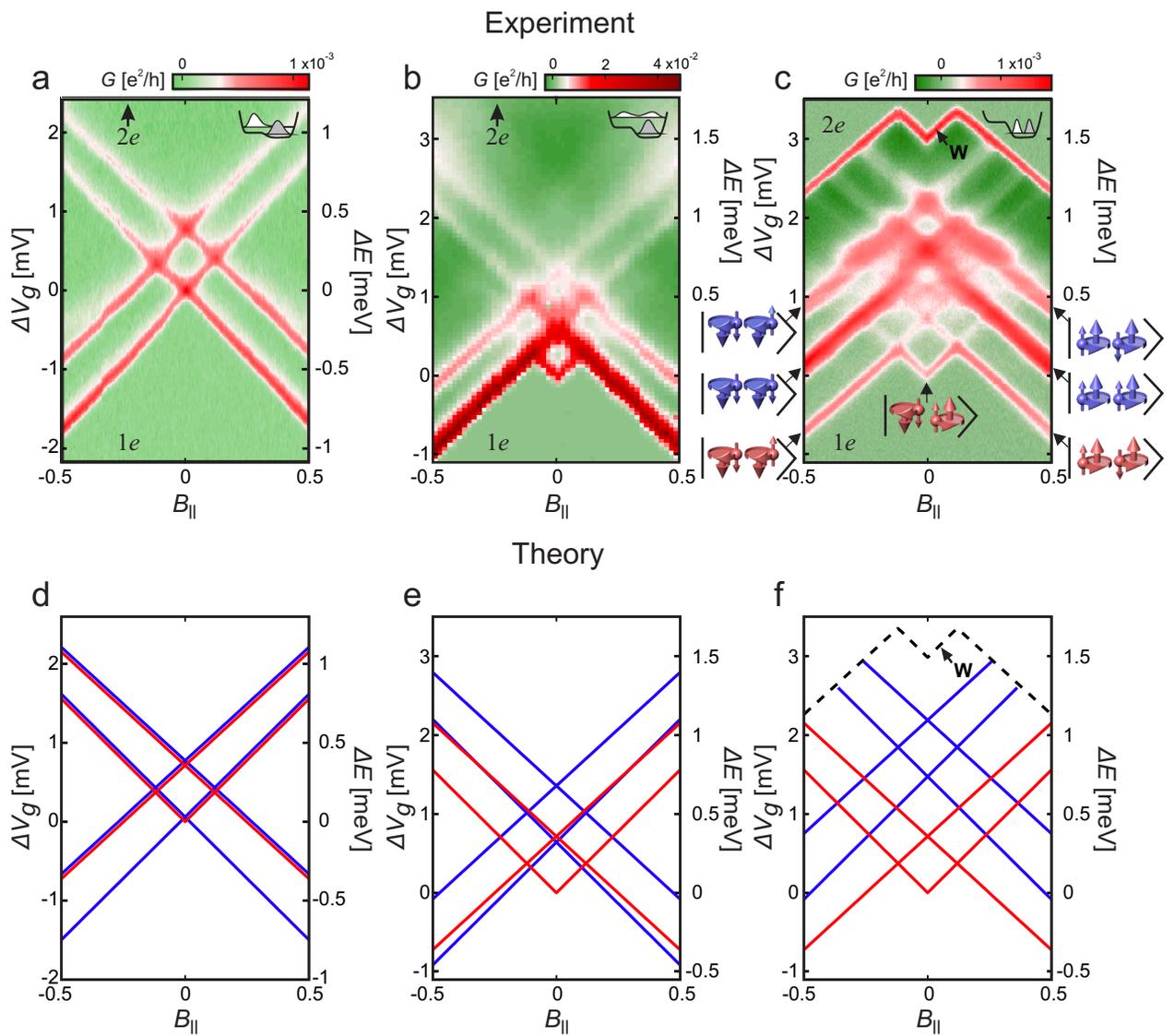

figure 3

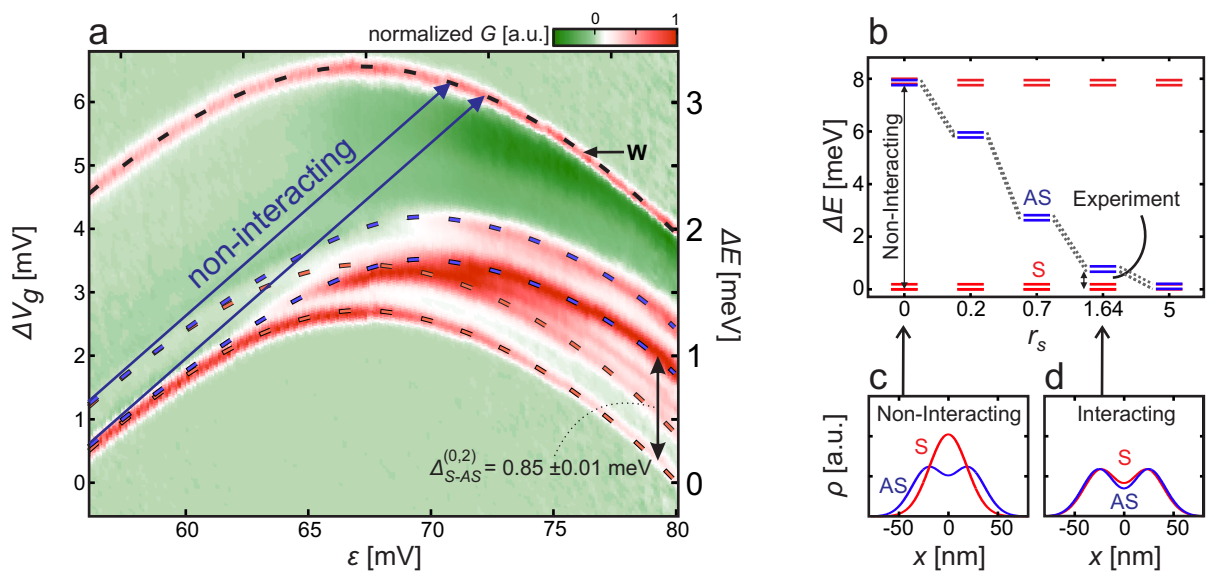

figure 4

# SUPPLEMENTARY INFORMATION

## Observation and Spectroscopy of a Two-Electron Wigner Molecule in an Ultra-Clean Carbon Nanotube

S. Pecker*[1], F. Kuemmeth*[2], A. Secchi[3,4£], M. Rontani[3], D. C. Ralph[5,6], P. L. McEuen[5,6] and S. Ilani[1]

[1]*Department of Condensed Matter Physics, Weizmann Institute of Science, Rehovot 76100, Israel*

[2]*Center for Quantum Devices, Niels Bohr Institute, University of Copenhagen, Universitetsparken 5, DK-2100 Copenhagen, Denmark*

[3]*CNR–NANO Reseach Center S3, Via Campi 213/a, 41125 Modena, Italy*

[4]*Department of Physics, University of Modena and Reggio Emilia, 41125 Modena, Italy*

[5]*Physics Department, Cornell University, Ithaca, New York 14853, USA*

[6]*Kavli Institute at Cornell, Cornell University, Ithaca, New York 14853, USA*

[£]*Present address: Institute for Molecules and Materials, Radboud University of Nijmegen, 6525 AJ Nijmegen, The Netherlands*

S1. **Symmetries of two-electron states in carbon nanotubes**

Electronic wavefunctions in carbon NTs consist of three components: a real-space component along the tube, a spin component, and an 'isospin' component related to the clockwise or counter-clockwise motion of the electron around the tube's circumference[†]. Although the isospin is formally also a spatial degree of freedom, energetically it is well decoupled from the spatial component along the tube and is in fact strongly coupled to the spin degree of freedom via spin-orbit coupling. Thus, the total wavefunction is naturally decomposed into a spatial and a spin-isospin components: $(spatial) \otimes (isospin \otimes spin)$. For a single electron, for each particle-in-a-box level in real space there are four possible spin-isospin combinations. Correspondingly, for two electrons there are $4 \times 4 = 16$ spin-isospin combinations. In this section we show how these states fall into multiplets with distinct symmetries and highlight which of these states should be visible in tunneling spectroscopy.

Since the total wavefunction of two electrons has to be anti-symmetric under exchange of electrons, the behavior in the spatial and the spin-isospin subspaces is anti-correlated: If the wave function is symmetric (S) in real-space it must be anti-symmetric (-) in spin-isospin space. Similarly, if it anti-symmetric (AS) in real-space, it has to be symmetric (+) in spin-isospin space. Note that throughout the main text we refer to the 2*e* states through their symmetry in real-space (S or AS), since this is the degree of freedom that naturally couples to Coulomb interactions.

The breakdown of the 16 two-electron spin-isospin combinations according to their symmetries is as follows: To form a 'singlet-like' (-) state in spin-isospin subspace we can combine a singlet in spin with a triplet in isospin and vice versa, yielding in total of $1 \times 3 + 3 \times 1 = 6$ states. To form a 'triplet-like' (+) state in spin-isospin subspace we can combine a spin triplet state with an isospin triplet state or a spin singlet state with an isospin singlet state, giving in total $3 \times 3 + 1 \times 1 = 10$ states.

---

[†] We note that both the spin and isospin degrees of freedom couple to a magnetic field, and in this work we use the terms "spin" and "isopin" to refer to the spin and isospin magnetic moments (which are the observables in our experiments) opposed to the angular momenta (which strictly speaking cease to be good quantum numbers in the presence of spin-orbit coupling and K-K' scattering).

To define more specifically these different states we denote the isospin of each electron by $K$ ($K'$) and its spin by $\downarrow$ ($\uparrow$), where the spin quantization is along the tube axis. One-electron states then read as $|K\uparrow\rangle$, $|K'\uparrow\rangle$, $|K\downarrow\rangle$, and $|K'\downarrow\rangle$, while the non-symmetrized two-electron states read as $|K\uparrow\rangle_1|K'\uparrow\rangle_2$, $|K'\uparrow\rangle_1|K'\uparrow\rangle_2$, etc., where the index refers to the first and second electron. After symmetrization or anti-symmetrization the relevant states become, for example:

$$|K\uparrow K'\uparrow\rangle^{\pm} \equiv \frac{1}{\sqrt{2}}(|K\uparrow\rangle_1|K'\uparrow\rangle_2 \pm |K'\uparrow\rangle_1|K\uparrow\rangle_2)$$

In the absence of spin-orbit coupling and electron-electron interactions all the states within the (+) or (-) multiplets are degenerate. However, in the presence of spin-orbit interactions the states split according to the relative alignment of their spin and isospin. For two electrons there are three different energetic configurations: If both electrons have parallel spin and isospin the energy is $-\Delta_{SO}$, if one has parallel alignment and another antiparallel alignment than the energy is zero, and if both are antiparallel than the energy is $\Delta_{SO}$ (where we assumed that the sign of the spin-orbit interactions prefers to parallel alignment of spin and isospin) (see Fig. 3 in Ref. 1).

In transport experiment that probe the $1e$ to $2e$ transition, only 7 out of the 16 states are visible due to selection rules imposed by sequential tunneling. Assuming that a single electron initially occupies the ground state $|K\downarrow\rangle$, a second electron can be added to any of the four spin-isospin states if the total wavefunction is spatially anti-symmetric, but can only occupy the other three spin-isospin states, if it is spatially symmetric, namely:

<div style="text-align:center">

Spatially Anti-Symmetric     Spatially Symmetric

$|K\downarrow\rangle + |K\downarrow\rangle \rightarrow |K\downarrow K\downarrow\rangle^{-}$

$|K\downarrow\rangle + |K'\uparrow\rangle \rightarrow |K\downarrow K'\uparrow\rangle^{-}$   or   $|K\downarrow K'\uparrow\rangle^{+}$

$|K\downarrow\rangle + |K\uparrow\rangle \rightarrow |K\downarrow K\uparrow\rangle^{-}$   or   $|K\downarrow K\uparrow\rangle^{-}$

$|K\downarrow\rangle + |K'\downarrow\rangle \rightarrow |K\downarrow K'\downarrow\rangle^{-}$   or   $|K\downarrow K'\downarrow\rangle^{-}$

</div>

In comparison, transitions such as $|K\downarrow\rangle \rightarrow |K\uparrow K\uparrow\rangle$ are not allowed, as they involve a change in the spin or isospin of both electrons simultaneously. Thus, if transport starts from one electron in the ground states then there are four anti-symmetric and three symmetric possible end states. The counting is summarized in table S1.

| Real-Space | Spin ⊗ Isospin | Total | Should appear in transport |
|---|---|---|---|
| Symmetric | - | 6 | 3 |
| Anti-Symmetric | + | 10 | 4 |

**Table S1** Counting the two-electron states according to their symmetries in real space.

S2. **Direct measurement of the single-particle level spacing**

In order to estimate the effect of interactions in our two-electron system, we compare in the main text the splitting between the symmetric and anti-symmetric multiplets, $\Delta_{S-AS}$, with the single-particle level spacing of the right dot, $\Delta_{ls}$. In this section we demonstrate how this level spacing is directly determined from the excitation spectrum of a single electron.

Figure S1a shows the conductance, $G$, measured as a function of gate voltage, $V_g$, and source-drain bias, $V_{sd}$, at the $(0,0) - (0,1)$ transition at $B_\parallel = 7$ T. This measurement directly reflects the energy spectrum of a single electron in the right dot. Three pairs of excitation lines, apparent as lines of high conductance, are visible: A low-energy pair (labeled $\alpha_1, \beta_1$) a pair at intermediate energies (labeled $\alpha_2, \beta_2$) and a pair at high energies, whose splitting is barely visible (labeled $\gamma_1, \delta_1$). The states $\alpha_1, \beta_1, \gamma_1, \delta_1$ can be directly associated with the four spin-isospin states of the lowest particle-in-a-box level. In fact, their energies perfectly fit the magnetic-field dependence of the states, shown as an orange double-cross in Fig. S1b: The lowest states correspond to $|K \downarrow\rangle$ and $|K \uparrow\rangle$ and their extracted splitting, $E_{\beta 1} - E_{\alpha 1} = 1.4$ meV, matches well the expected splitting: $E_{K\uparrow} - E_{K\downarrow} = \Delta_{SO} + 2\mu_s B = 1.24$ meV ($\mu_s$ is the spin contribution to the magnetic moment). States of opposite isospin are split by the magnetic field by $E_{K'} - E_K = 2\mu_{orb} B \approx 15$ meV ($\mu_{orb}$ is the orbital contribution to the magnetic moment) matching the observed splitting between the $(\alpha_1, \beta_1)$ pair and the $(\gamma_1, \delta_1)$ pair. Finally, the high-energy pair of $K'$ states is expected to be split by $E_{K'\uparrow} - E_{K'\downarrow} = 2\mu_s B - \Delta_{SO} = 0.5$ meV, also faintly observed in the measurement. This clearly demonstrates that the four excitations $\alpha_1, \beta_1, \gamma_1, \delta_1$ are the four spin-isospin states of a single electron in the lowest particle in a box level.

The remaining pair of lines at intermediate energies $(\alpha_2, \beta_2)$ are separated from the low-energy pair by $E_2 - E_1 = \frac{1}{2}(E_{\alpha 2} + E_{\beta 2}) - \frac{1}{2}(E_{\alpha 1} + E_{\beta 1}) = 7.8$ meV $< 15$ meV, and so their isospin must be parallel to the field $(K)$. This pair must then correspond to the first excited level, which for a single electron corresponds to the second particle-in-a-box state. Its corresponding magnetic field dependence is shown in purple in Fig S1b. Specifically we see that the splitting between the states $\alpha_2, \beta_2$, $E_{\beta 2} - E_{\alpha 2} = 1.2$ meV fits well the predicted one $E_{K\uparrow} - E_{K\downarrow} = 1.24$ meV.

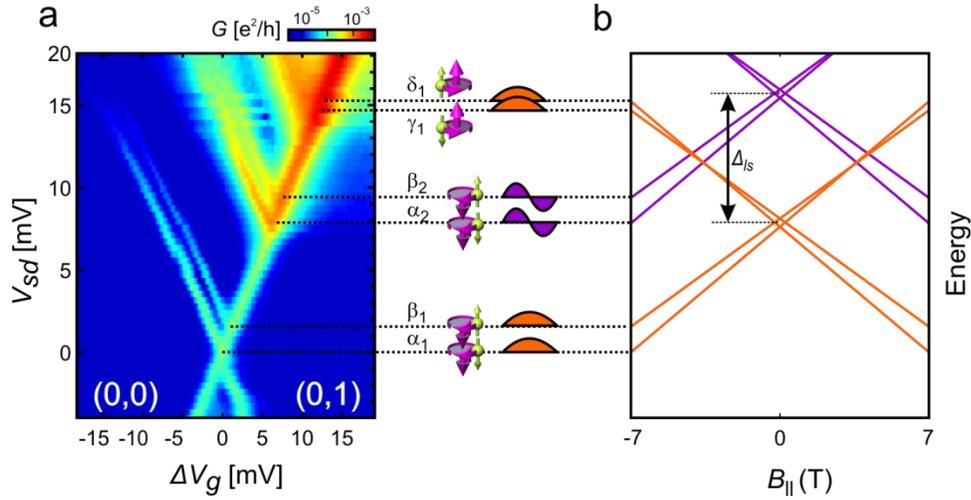

**Figure S1: Measurement of the single particle level spacing of the right dot. a**. One-electron excitation spectrum at the transition (0,0)-(0,1). Conductance $G$ is plotted as a function of gate voltage $\Delta V_g$ and source-drain bias $V_{sd}$ at magnetic field $B_\parallel = 7$ T featuring parallel lines corresponding to *1e* excitations. **b**. Predicted magnetic-field dependence of the *1e* spectrum allowing identification of excitation in **a**. Spin (green) and isospin (magenta) are illustrated by up- and down-pointing arrows. States of the first (second) particle-in-a-box state appear in orange (purple).

Having identified all the states, we can unambiguously determine the single-particle level spacing of the right dot from the energy difference between the $(\alpha_2, \beta_2)$ and $(\alpha_1, \beta_1)$ pairs. The level spacing is thus $\Delta_{ls} = 7.8$ meV.

### S3. Comparison of one-electron and two-electron excitation spectra.

In the main text we classify the *2e* states by their symmetry properties according to magnetic-field fingerprints based on the single-particle magnetic properties (Fig. 3). In this section we quantitatively compare the *1e* spectrum (Fig. S2a) reported before[2] and the *2e* spectrum at the transition to the (1,1) configuration (Fig. S2b, duplicating Fig. 3a in the main text). The two spectra are found to be practically identical, supporting the use of single-particle magnetic properties in the main text.

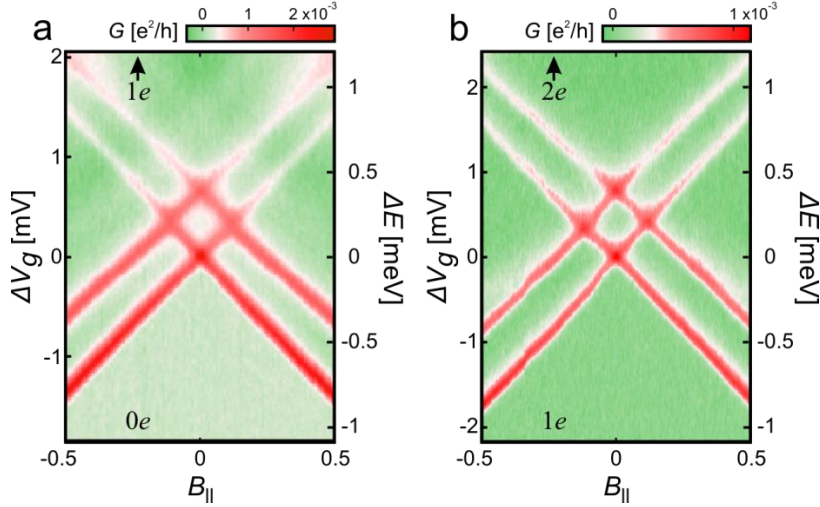

**Figure S2: Comparison of one-electron and two-electron excitation spectra. a,** One-electron and **b,** Two-electron magnetic-field dependence of the excitation spectrum: $G$ is plotted as a function of $\Delta V_g$ and $B_\parallel$ at finite bias ($V_{sd} = -2.2$ mV in panel a, $2$ mV in panel b). Lines of enhanced conductance correspond to tunneling via *1e* (a) and *2e* (b) states. Panel b reproduces Fig. 3a of the main text.

We start by comparing the magnetic moments in the two spectra. From the *2e* spectrum we extract $\mu = \pm 1.61$ meV/T and $\mu = \pm 1.48$ meV/T for the lower and upper crosses respectively. The average moment corresponds to the orbital contribution to the moment, $\mu_{orb} = 1.55 \pm 0.01$, whereas the difference corresponds to Zeeman splitting, giving the gyromagnetic factor $g = 2.2 \pm 0.4$, both in good agreement with *1e* moments reported before ($\mu_{orb} = 1.55$, $g = 2.1 \pm 0.1$). This allows us to attribute the splitting at $B_\parallel = 0$ to spin-orbit coupling, which measures to be $\Delta_{so} = 0.36 \pm 0.02$ meV, in excellent agreement with the *1e* spin-orbit coupling $\Delta_{so} = 0.37 \pm 0.02$ meV.

## S4. **Inter-valley Coulomb interactions in the observed excitation spectrum**

The measured excitation spectra of two electrons (Fig. 3c) or two holes (Fig. S4a) in a single dot, feature two more lines in addition to the seven states described in the main text. In this section we show how interactions, combined with long-lived excited states, may lead to the two extra lines that are not captured by the non-interacting picture. This section starts by describing the effect of inter-valley Coulomb interactions on the complete two-electron excitation spectrum. We then describe how all the lines in the measured spectra can be accounted for by non-equilibrium transport.

We define as inter-valley backscattering (VBS) the Coulomb-interaction process that exchanges the isospins of two interacting electrons (or holes) having opposite isospins. We first note that the effect of VBS is generally small with respect to forward scattering, i.e. Coulomb-interaction processes which do not involve valley exchange, such as inter-valley Hartree-like interactions and all kinds of intra-valley interactions. This allows us to treat VBS within first-order perturbation theory. Forward scattering and quantum confinement then determine the energy splitting between the symmetric and anti-symmetric multiplets, as described in the main text, whereas VBS and spin-orbit interaction determine the splitting within each multiplet, as we describe below. Second, qualitatively speaking, the effect of VBS scattering is short ranged, since it involves large momentum transfer between the scattered particles. It therefore has negligible effect on the spatially anti-symmetric multiplet, where the two electrons have small probability to be one close to the other along the nanotube axis.

By including the effect of VBS in our exact-diagonalization calculation we obtain the *2e* spatially symmetric spectrum presented in Fig. S3b. The calculation shows two main differences compared to the non-interacting spectrum (Fig. S3a): First, the four-fold degeneracy of the central line at $B_\parallel = 0$ is broken by $\Delta_{VBS}$. Second, the apparent spin-orbit gap is enhanced $\Delta_{SO} \rightarrow \Delta_{SO}^* = \sqrt{\Delta_{SO}^2 + \Delta_{VBS}^2}$. The magnetic fingerprint of the symmetric multiplet is therefore slightly altered by interactions.

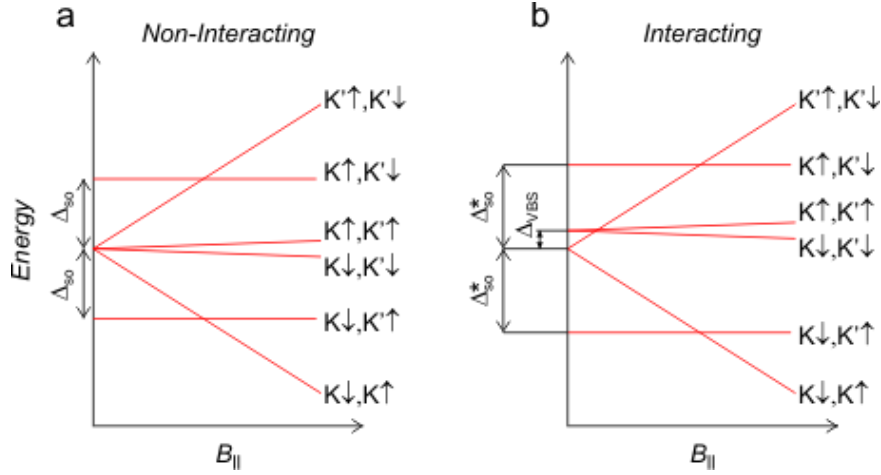

**Figure S3 Predicted effect of inter-valley backward scattering on the spatially symmetric *2e* excitation spectrum. a.** Schematic energy spectrum of symmetric *2e* states as a function of $B_\parallel$ in the non-interacting picture. The states are split at $B_\parallel = 0$ by spin-orbit coupling $\Delta_{so}$. **b.** same as **a** in the presence of inter-valley backward scattering. The four-fold degeneracy as $B_\parallel = 0$ is lifted by $\Delta_{VBS}$ and $\Delta_{so}$ is replaced by $\Delta_{so}^* = \sqrt{\Delta_{so}^2 + \Delta_{VBS}^2}$.

Whereas VBS leads to energy shifts in the *2e* spectrum, interactions alone do not fully account for the measured spectrum. As long as transport starts and ends with a single electron occupying the ground state of the dot, as described in the main text, it leads to 7 spectroscopic lines. However, additional lines would appear if the starting electron occupied instead a *metastable* state having a long lifetime. A natural candidate for such a metastable state is $|K' \uparrow\rangle$, which has spin *and* isospin opposite to the ground state $|K \uparrow\rangle$ leading to long expected relaxation times. Table S2 lists all the processes starting in the $|K' \uparrow\rangle$ state and ending in spatially symmetric *2e* states, along with their corresponding addition energies in both the non-interacting and interacting cases. The table enumerates three additional *2e* states that are now accessible, two of which (lines 4-5 in the table) result in extra lines in the spectrum, whereas the last one does not appear in the transport as it lies in the Coulomb-blockaded region (addition energies listed in Table S2 should not be confused with 2e energies plotted in Fig. S3).

The measured spectra fully agree with the above picture and from them we extract $\Delta_{BS} = 0.21 \pm 0.01$ meV for the *2e* molecule, in agreement with the enhanced $\Delta_{SO}^* = 0.40$ meV, and $\Delta_{BS} = 0.19 \pm 0.01$ meV for the *2h* molecule, in agreement with $\Delta_{SO}^* = 0.26$ meV. It is interesting to note, however, that whereas the calculations agree qualitatively with the data and its symmetries, they yield $\Delta_{BS}$ smaller by an order of magnitude and opposite in sign compared to the value extracted from the experiment. This remaining puzzle is a challenge for future theories trying to understand the finer effects of interactions in this system.

| # | Initial *1e* state | Final *2e* state | Addition energy | | Appears in transport? |
|---|---|---|---|---|---|
| | | | Non-Interacting | VBS induced shift | |
| 1 | $\|K\downarrow\rangle$ (ground state) | $\|K\downarrow K'\uparrow\rangle_S$ | $E_{K'\uparrow}$ | $\Delta_{SO} - \Delta_{SO}^*$ | Yes |
| 2 | | $\|K\downarrow K\uparrow\rangle_S$ | $E_{K\uparrow}$ | 0 | Yes |
| 3 | | $\|K\downarrow K'\downarrow\rangle_S$ | $E_{K'\downarrow}$ | $\Delta_{VBS}$ | Yes |
| 4 | $\|K'\uparrow\rangle$ (metastable) | $\|K'\uparrow K\uparrow\rangle_S$ | $E_{K\uparrow}$ | $\Delta_{VBS}$ | Yes |
| 5 | | $\|K'\uparrow K'\downarrow\rangle_S$ | $E_{K'\downarrow}$ | 0 | Yes |
| 6 | | $\|K'\uparrow K\downarrow\rangle_S$ | $E_{K\downarrow}$ | $\Delta_{SO} - \Delta_{SO}^*$ | No (In blockaded region) |

**Table S2 Spectroscopic lines of the spatially symmetric multiplet taking into account the metastable state $|K'\uparrow\rangle$ and backward scattering.** Lines 1-3 correspond to the three lines discussed in the main text. Lines 4-6 correspond to three additional lines. Lines 1-5 lie within the spectroscopic window and thus appear in transport.

S5. **Excitation spectrum and detuning dependence of two holes in a single dot.**

In the main text we present spectroscopic evidence for the formation of a 2*e* Wigner-molecule state. In this section we present the detuning dependence and magnetic-field dependence of 2*h* excitations, both remarkably similar in all aspects to the 2*e* data. This demonstrates that all the observations presented for electrons in the main paper are in fact generic and do not depend on details such as the charge of the carriers, disorder, and the strength or sign of spin-orbit coupling.

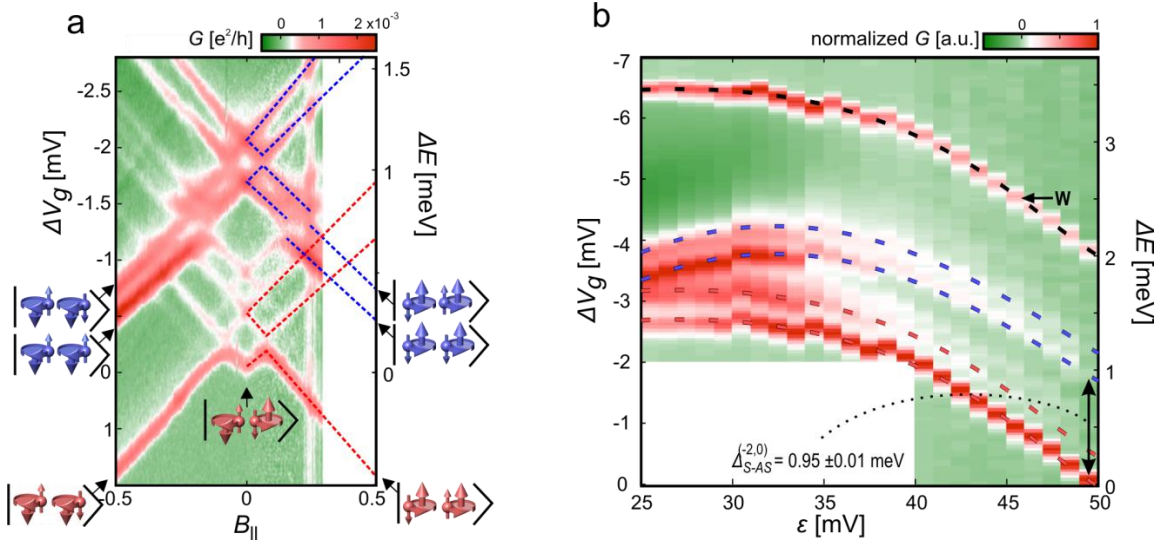

**Figure S4: Excitation spectrum of a two-hole Wigner molecule. a.** Magnetic-field depedent spectrum in the (2h,0) configuration. $G$ is plotted as a function of $B_\parallel$ and $V_g$. ($V_{sd} = -2.5$ mV, $\epsilon = 50$ mV) **b**. Detuning dependence of the spectrum at $B_\parallel = 0$ and $V_{sd} = 2$ mV. Dashed lines in **a** and **b** – guides to the eye following the spatially anti-symmetric (blue) and symmetric (red) multiplets and the spectroscopic window (black, W). The inter-multiplet splitting $\Delta_{S-AS}$ is extracted from **a**.

Figure S4a presents the magnetic-field dependent *2h* excitation spectrum at the transition $(1h, 0) \to (2h, 0)$. The conductance measured with $V_{sd} = -2.5$ mV is plotted as a function of $B_\parallel$ and $V_g$ (converted to energy on the right y-axis). The lines are matched with the magnetic-field fingerprints: The cusp at $B_\parallel = 0$ and the cross above it are identified with the *2h* symmetric multiplet, whereas the top-most double-cross is identified with the anti-symmetric multiplet. The remaining line is due to inter-valley interactions, as described in section S4 above. The data compares very well to the *1h* magnetic-field dependence reported before[2]: The slopes match the *1h* magnetic moments, and the $B_\parallel = 0$ splitting agrees in sign and magnitude with the *1h* spin-orbit coupling ($\Delta_{so} = 0.17 \pm 0.02$ meV in the symmetric multiplet and $\Delta_{so} = 0.20 \pm 0.01$ meV in the anti-symmetric). Finally, the avoided crossing between the $|K \uparrow K \downarrow\rangle$ and $|K \uparrow K' \downarrow\rangle$ states seen in both multiplets at $B_\parallel = \pm 0.1$ T matches the *1h* disorder-induced valley mixing ($\Delta_{KK'} = 0.1$ meV).

Figure S4b presents the detuning-dependent excitation spectrum at the *1h-2h* transition. The conductance is plotted as a function of $V_g$ (converted to energy on the right y-axis) and $\epsilon$ at $B_\parallel = 0$ and $V_{sd} = 2$ meV. Five lines are seen within the spectroscopic window (labeled W). From their magnetic-field dependence we identify the two bottom-most lines as the symmetric multiplet and the two top-most lines as the anti-symmetric multiplet, while the remaining line is a result of inter-valley interactions (see S4). Both multiplets undergo a transition from the (1$h$,1$h$) charge configuration (up-going slope) to the (2$h$,0) configuration (down-going slope) as a function of detuning. The 2$h$ spectrum presented in Fig. S4a (measured at $\epsilon = 50$ meV) therefore corresponds to two holes in a single dot. Similar to the (0,2$e$) configuration, the multiplet splitting in the (2$h$,0) configuration, measured to be $\Delta_{S-AS}^{(2h,0)} = 0.95 \pm 0.01$ meV, is quenched by an order of magnitude with respect to the single-particle level spacing measured in the (1$h$,0) configuration[2] $\Delta_{ls}^{(1h,0)} = 11$ meV.

We conclude that the 2$h$ system presents quenching of the inter-multiplet splitting, similar to that of the 2$e$ system, even though the two systems differ considerably in spin-orbit coupling (holes feature half as strong coupling of opposite sign compared to electrons, favoring anti-parallel spin and isospin compared to parallel in electrons) and in disorder-induced valley mixing (holes feature twice as strong mixing), have opposite charges, and sit in different dots. This supports that the formation of a Wigner-molecule state, manifested by the inter-multiplet excitation energy quenching, is indeed a generic phenomenon which does not depend on the above details of the system.

S6. **The complete detuning dependence of excited-state spectra from positive and negative bias measurements.**

Due to asymmetric coupling of the double dot to the two leads, excitations may appear stronger or weaker depending on the bias direction. In general, transport via excited states that have a strong tunneling barrier to the drain (source) is more visible for positive (negative) bias. Therefore the process (0,1) → (1,1) → (0,1) is more visible in positive bias (Fig. 1c) while (0,1) → (0,2) → (0,1) is more visible in negative bias (Fig. 1d). The detuning dependent spectrum presented in the main text (Fig. S5a, duplicating Fig. 4a in the main text) was measured at negative bias, and its low-detuning excitations are thus very faint. However, the missing lines appear clearly at positive bias (Fig. S5b), and the spectrum is seen to continuously evolve with detuning. This is demonstrated by plotting the same dashed guidelines on top of the two spectra, taking into account gating effects induced by the opposite source-drain bias.

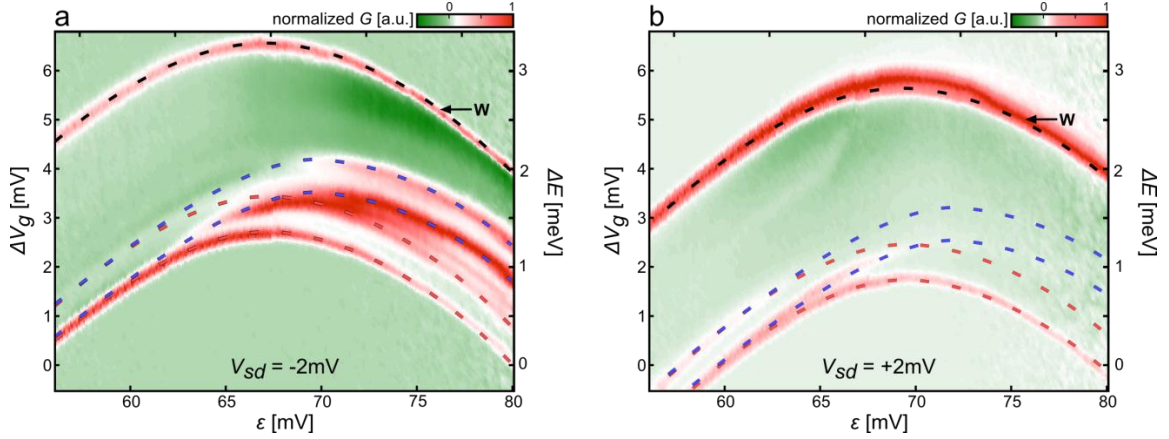

**Figure S5: Complete detuning dependence of the *2e* excitation spectrum. a.** Measurement at negative bias $V_{sd} = -2$ **mV** (duplicating Fig. 4a in the main text) $G$ is plotted as a function of $\epsilon$ and $V_g$ (converted to energy on the right y-axis). **b.** same as **a** with $V_{sd} = +2$ **mV** showing the low-detuning part of the spectrum that is missing on a. Identical guidelines are plotted on top of the two measured spectra, following the symmetric (red) and anti-symmetric (blue) multiplets and the spectroscopic window (black, W). The guidelines are shifted by $\Delta V_g = 1$ **mV** and $\Delta \epsilon = 2.4$ **mV** due to gating by the source-drain bias.

## S7.  **Exact diagonalization**

In the main text we show the evolution with $r_s$ of two-electron excitation energies in a NT quantum dot (Fig. 4b), together with selected charge-density profiles (Figs. 4c and 4d). These results are obtained by means of the exact-diagonalization method, also known as full configuration interaction. In this section we briefly review the key steps of the calculation. An overview of the method is reported in Ref. 3 and full details on its application to carbon NTs are provided in Ref. 4. With respect to Ref. 4, here we have improved our treatment of VBS interaction.

Since the quantum-dot confinement potential is soft, being induced by electric gates, its generic low-energy dependence on the NT-axis coordinate $x$ is quadratic. Therefore, within the envelope-function and effective-mass approximation, the one-electron wavefunction $\Psi_{n\tau\sigma}(\vec{r},s)$ may be written as

$$\Psi_{n\tau\sigma}(\vec{r},s) = AF_n(x)\psi_\tau(\vec{r}) \otimes \chi_\sigma(s) = \varphi_{n\tau}(\vec{r}) \otimes \chi_\sigma(s),$$

where $A$ is a normalization constant, $F_n(x)$ is the envelope-function $n$th eigenstate of the one-dimensional harmonic oscillator ($n = 0, 1, 2, \ldots$), slowly varying with respect to the graphene lattice constant, $\chi_\sigma(s)$ is the electron spinor (with $\sigma = \pm 1$ being equal to the third component of spin, in units of $\hbar/2$), and $\psi_\tau(\vec{r})$ is the bulk Bloch state whose wave vector is located at the bottom of conduction-band valley $\tau$ in reciprocal space (valley $K$ if $\tau = +1$, valley $K'$ if $\tau = -1$, the quantum number $\tau$ being the isospin). The one-electron energy $E_{n\tau\sigma}$ is

$$E_{n\tau\sigma} = \hbar\omega_0(n + 1/2) + \Delta_{so}\tau\sigma/2 + \mu_s B\sigma - \mu_{orb} B\tau,$$

where the harmonic-oscillator energy quantum $\hbar\omega_0 = 7.8$ meV is equal to the observed level spacing $\Delta_{ls}$.

Using $\Psi_{n\tau\sigma}(\vec{r}, s)$ as one-electron basis set, the interacting Hamiltonian $\hat{H}$ in second-quantization acquires the form

$$\hat{H} = \hat{H}_{SP} + \hat{V}_{FW} + \hat{V}_{BW}.$$

The single-particle term $\hat{H}_{SP}$ is

$$\hat{H}_{SP} = \sum_{n\tau\sigma} E_{n\tau\sigma} \hat{c}^+_{n\tau\sigma} \hat{c}_{n\tau\sigma},$$

where the fermionic operator $\hat{c}^+_{n\tau\sigma}$ ($\hat{c}_{n\tau\sigma}$) creates (destroys) an electron with spin σ and isospin τ in the $n$th level of the harmonic oscillator. The forward-scattering (FS) term

$$\hat{V}_{FS} = \frac{1}{2} \sum_{nn'n''n'''} \sum_{\tau\tau'\sigma\sigma'} V^{\tau\tau'\tau'\tau}_{nn'n''n'''} \hat{c}^+_{n\tau\sigma} \hat{c}^+_{n'\tau'\sigma'} \hat{c}_{n''\tau'\sigma'} \hat{c}_{n'''\tau\sigma}$$

includes all intra-valley scattering processes due to Coulomb interaction as well as Hartree-like inter-valley scattering terms. The FS matrix element

$$V^{\tau\tau'\tau'\tau}_{nn'n''n'''} = \iint d\vec{r} d\vec{r}' \varphi^*_{n\tau}(\vec{r}) \varphi^*_{n'\tau'}(\vec{r}') U(\vec{r} - \vec{r}') \varphi_{n''\tau'}(\vec{r}') \varphi_{n'''\tau}(\vec{r})$$

depends on the static dielectric constant $\kappa_r$ of the electrostatic environment, here treated as a free parameter, through the so-called Ohno potential,

$$U(\vec{r}-\vec{r}\,') = \frac{e^2}{\sqrt{e^4/U_0^2 + \kappa_r^2|\vec{r}-\vec{r}\,'|^2}},$$

which interpolates the two limits of Coulomb-like long range and Hubbard-like short range interactions (with Hubbard $p_z$-site parameter $U_0 = 15$ eV). The six-dimensional integral $V_{nn'n''n'''}^{\tau\tau'\tau'\tau}$ is evaluated by neglecting the overlap of $p_z$ orbitals on different sites and using the slow variation in space of the envelope function $F_n(x)$.

Backward interactions are included in the term

$$\hat{V}_{VBS} = \frac{1}{2}\sum_{nn'n''n'''}\sum_{\tau\sigma\sigma'} V_{nn'n''n'''}^{\tau-\tau\tau-\tau}\, \hat{c}_{n\tau\sigma}^+\, \hat{c}_{n'-\tau\sigma'}^+\, \hat{c}_{n''\tau\sigma'}\, \hat{c}_{n'''-\tau\sigma},$$

which exchanges the valleys of two scattering electrons when they have opposite isospins, otherwise it has no effect. The evaluation of VBS matrix elements

$$V_{nn'n''n'''}^{\tau-\tau\tau-\tau} = \iint d\vec{r}d\vec{r}\,'\varphi_{n\tau}^*(\vec{r})\varphi_{n'-\tau}^*(\vec{r}\,')U(\vec{r}-\vec{r}\,')\varphi_{n''\tau}(\vec{r}\,')\varphi_{n'''-\tau}(\vec{r})$$

lies outside the range of applicability of the standard envelope-function theory. We will show elsewhere that such matrix elements contain the short-range part of interaction, weakly depend on the unknown NT chirality, and are smaller by orders of magnitude than FS matrix elements. On the other hand, FS matrix elements depend only on macroscopic NT parameters such as the radius $R$, which is deduced by the measured value of $\mu_{orb}$.

We exactly diagonalize the FS Hamiltonian

$$\hat{H}_{SP} + \hat{V}_{FS},$$

which is a matrix in the Fock space of Slater determinants $|\Phi_i>$ that are obtained by filling with two electrons in all possible ways the lowest 50 one-electron orbitals $\varphi_{n\tau}(\mathbf{r})$ (two-fold spin degenerate at $B = 0$). Both ground and excited two-electron states, $|\Psi_n>$, expanded on the basis of Slater determinants,

$$|\Psi_n\rangle = \sum_i c_i^n |\Phi_i\rangle,$$

are obtained numerically, together with their energies, by means of the parallel home-built code *DonRodrigo*. The diagonalization proceeds in each Hilbert-space sector labeled by the total spin component along the NT axis, total isospin, and parity under mirror reflection with respect to a plane perpendicular to the NT axis, placed in the middle of the quantum dot. The effect of VBS terms on eigenstates $|\Psi_n>$ of FS Hamiltonian is considered at the level of first-order degenerate perturbation theory.

The code output (i.e., the expansion coefficients $c_i$) is post-processed in order to obtain the charge density $\rho(x)$ for the $n$th excited state at given $r_s$,

$$\rho(x) = \sum_{ij} c_i^n c_j^{n*} \langle \Phi_j | \sum_{mm'\tau\tau'\sigma} F_m^*(x) F_{m'}(x) \hat{c}_{m\tau\sigma}^+ \hat{c}_{m'\tau'\sigma} | \Phi_i \rangle.$$

The density parameter $r_s$ is estimated as the ratio of the characteristic harmonic oscillator length to the effective Bohr radius $a_B^*$,

$$r_s = \left(\frac{\hbar}{m^*\omega_0}\right)^{1/2} \left(\frac{\hbar^2 \kappa_r}{m^* e^2}\right)^{-1} = \frac{2e^2 (m^*)^{1/2}}{\hbar^{3/2} (\omega_0)^{1/2} \kappa_r},$$

where the electron effective mass $m^*$ is obtained through the formula

$$m^* = \frac{\hbar^2}{3R\gamma},$$

with γ = 0.54 eV nm being the graphene π-band parameter.